\documentclass[12pt]{article}

\usepackage{setspace}
\usepackage{natbib}
\usepackage{times,amsmath,float}
\usepackage{amssymb}
\usepackage{graphicx}
\usepackage{subfigure}
\usepackage{amsfonts}
\usepackage{epsfig}
\usepackage{anysize}
\usepackage{xtab}
\usepackage{lscape}
\usepackage{geometry}
\usepackage{longtable}
\usepackage{appendix}
\usepackage{setspace}
\usepackage{epstopdf}
\usepackage{chngcntr}
\usepackage{booktabs}
\usepackage{multirow}
\usepackage{caption,booktabs,array}
\usepackage{float}
\usepackage{adjustbox}
\usepackage{url}
\usepackage{multicol}
\usepackage{placeins}
\usepackage{pdfpages}
\usepackage{multicol}
\usepackage{titling}
\marginsize{2.5cm}{2.5cm}{2.5cm}{2.5cm}

\begin{document}

 \setlength{\parindent}{2.1em}
\doublespacing 

\begin{center}

\vspace{10cm}
\Large Reading Macroeconomics From the Yield Curve: \\
The Turkish Case
\vspace{8cm}

Ipek Turker \& Bayram Cakir\\
\vspace{6cm}
2019
\end{center}

\newpage
\begin{center}
ABSTRACT\\
Reading Macroeconomics From the Yield Curve: The Turkish Case\\
\end{center}
\textbf{ } \\

\noindent 
This paper aims to analyze the relationship between yield curve -being a line of the interests in various maturities at a given time- and GDP growth in Turkey. The paper focuses on analyzing the yield curve in relation to its predictive power on Turkish macroeconomic dynamics using the linear regression model.  To do so, the interest rate spreads of different maturities are used as a proxy of the yield curve. Findings of the OLS regression are similar to that found in the literature and supports the positive relation between slope of yield curve and GDP growth in Turkey. Moreover, the predicted values of the GDP growth from interest rate spread closely follow the actual GDP growth in Turkey, indicating its predictive power on the economic activity.

\newpage

\doublespacing
\section{Introduction}

Yield curve is a line of the interest rates in various maturities at a given time. For example, US yield curve shows the yearly interest rate for maturity rates of 3 months, 2 years, 5 years and 30 years. Looking at the interests rate of different maturities can help to read macroeconomics as it has a predictive power on future level of economic activities and output. Slope of the yield curve, which indicates the difference of interest rates, is a simple but strong indicator of macroeconomic conditions because it shows the difference in long term and short term interest rates. The market short term and long term interest rates can affect the investors' decisions. For instance, the investors will tend to invest more in real economic activities when short term interest rate is low and expect an increase in interest rate in the future (as long term interest rate is high and they do not want to be caught in lower interest rates in the future). Therefore, there is a wide literature which tries to estimate future macroeconomic conditions of the economies using the interest rate spreads.\\

There are three types of yield curves in terms of the sign of their slopes where each type can signal different state of the future of the economy:\\

The first one is called normal yield curve which has the positive slope which refers to higher interest rate level for longer term maturity than shorter term maturities (Figure 1). 

\begin{figure}[H]
\centering
\includegraphics[scale=1]{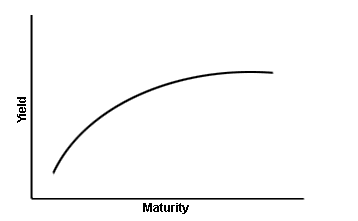}
\caption{Normal Yield Curve}
\end{figure}

The signal from the normal yield curve is that interests rates may continue to increase more as market expectations of the future interest rates is in an up trend. As a result, investors will look to invest their capital more on stocks and other financial instruments  related to real economic activities hoping that they will get higher returns from risk free investment in the future. Low rate of short interest rate will also cause money to be invested in real economic activities. This will create a downward pressure in the short term interest rates which will make the yield curve more steeper. Therefore, the increase in real economic activities will cause a economic expansion and more economic output as money flows to the real sector.\\

\newpage
The second type of yield curve is called inverted yield curve which has a negative slope with lower interest rate level as maturity increases (Figure 2).

\begin{figure}[H]
\centering
\includegraphics[scale=1]{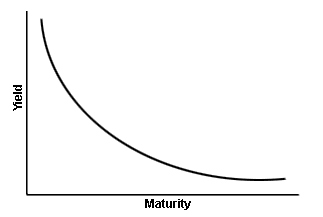}
\caption{Inverted Yield Curve}
\end{figure}

This yield curve is the least common one, and it is associated with the slower future economic output growth. The signal is exactly reverse of Normal Yield Curve because the market expectations of future interest rate is lower and in a downward trend. As a result, investors will invest in risk free bonds as they expect the future interest rates to decline more. Thus, the economic activities and output growth gets lower due to less investment in real economic activities. Moreover, negative slope of yield curve is a signal of lower interest rates in future as a result of possible government intervention to crisis. \\

The third yield curve is called a flat yield curve which indicates that the interest rate is similar for long and short term maturities. A flat yield curve is a signal of transition in the economy. If the economy is going to a recession from an expansion period, long term interest rates tend to fall and short term interest rates tend to increase. This makes the normal yield to curve to flat yield curve. If the economy is going to an expansion period from a recession, short term interest rates tend to rise and long term interest rates tend to fall, which result in a flat yield curve that comes from an inverted yield curve. \\

Following this economic intuition, there is a wide literature that uses the difference of interest rates for various maturities, analyzing the relationship between interest rate spread and macroeconomic conditions.

\section{Relevant Studies Regarding Yield Curve and Economic Performance} 

Clark (1996) refers to the yield curve as a "near-perfect tool for economic forecasting". Moreover, yield curve includes much more information than what is expected. Future inflation or future monetary policy implications are just some of the information that can be extracted from the yield curve. On the light of this idea, there are significantly many studies that try to extract information from spread of interest rates to forecast economic activities. Positive spread would indicate a normal yield curve and negative spread indicates an inverted yield curve. This section will be presenting the literature and findings of them.\\

\subsection{Yield Curve as a Predictor of GDP Growth}
Firstly, Joseph. et al. (1996) used the spread of interest rates between 10 years and 3 months to forecast the GDP growth for their data from 1961 to 1995 after their observation of positive correlation between these two. In the figure below, the 4 quarter GDP growth and lagged value of interest rate spread between 10 year and 3 months interest rate is plotted and they look to move in the same direction. 

\begin{figure}[H]
\centering
\includegraphics[scale=1]{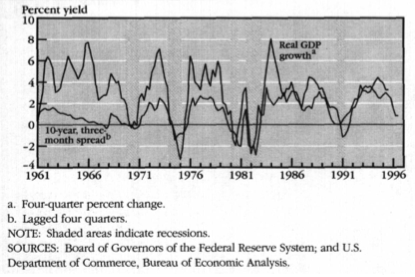}
\caption{GDP Growth and Lagged Interest Rate Spread}
\end{figure}

They run the following regression to see if interest rate forecast the GDP growth where coefficient of interest rate would be significant.\\

$\frac{RGDP_{t+4}-RGDP_t}{RGDP_T}=\alpha+\beta spread_t+\epsilon_t$\\

The results are the following:\\

\includegraphics[scale=0.5]{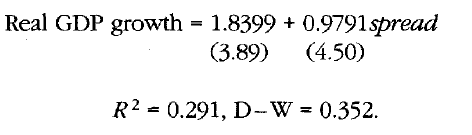}

This regression results suggests that 1 percent increase in the interest rate spread leads to around 1 percent increase in the GDP growth as the coefficient is 1 percent. This suggests that steeper yield curve reflects better future economic performance in the economy. Moreover, $R^2$ represents the explanatory power of interest rate spread on GDP growth. In the next figure, actual GDP growth and predicted GDP growth is plotted, forecasting the ability of interest rate spread. \\

\begin{figure}[H]
\centering
\includegraphics[scale=0.7]{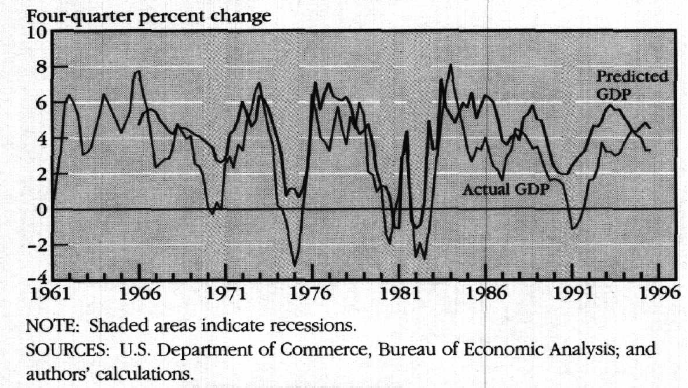}
\caption{Actual and Predicted GDP}
\end{figure}

\vspace{0.5cm}

Secondly, Harvey (1989) also conducted as similar analysis by using different interest rates for the independent variable. The spread variable is the difference between 10 years interest rate and 5 years interest rate. The dependent variable to forecast is GNP rather than GDP. The data of this paper starts in 1954 and ends in 1989.  The paper runs the following regression:\\

$$Growth_{t+1:t+5}= a+b(Yield Spread)_t+u_{t+5}$$\\

The coefficient b refers to the average risk tolerance in the economy. This analysis also shows the position relationship between Yield Spread and Growth rate as b coefficient is positive and significant. Out of sample analysis of this regression shows that interest rate spread can explain 35\% of the variation in GNP growth. The next figure shows the successful performance of the predicted GNP growth from the model as the predicted GNP growth moves closely to actual GNP growth.\\ 

\begin{figure}[H]
\centering
\includegraphics[scale=0.7]{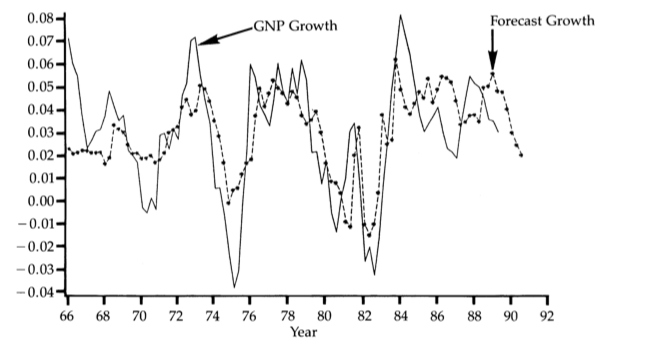}
\caption{Actual and Predicted GNP}
\end{figure}

Moreover, this study also compares the forecasting performance of yield curve on forecasting with other financial variables such as Stock Returns, lagged values of GNP. The comparison also suggests that the explanatory power of yield curve on GNP is higher than the other financial indicators. \\

A more comprehensive study was done by Dotsey (1998) by comparing the results for different time horizon forecasting. In this study, by using the difference in 3 months and 10 years interest rates between the periods 1955 and 1997, following regression is run:\\

$$(\frac{400}{k})\ln{\frac{y{t+1}}{y_t}}=\alpha_0+\alpha_1 s_t+\epsilon_t$$\\

This regression is run for different time horizon forecasting of 2, 4, 6 and 8 years in different periods to see if forecasting performance decrease over time. The following figure shows the results of the regression coefficient $\alpha_1$ for different sample periods:\\

\begin{figure}[H]
\centering
\includegraphics[scale=0.6]{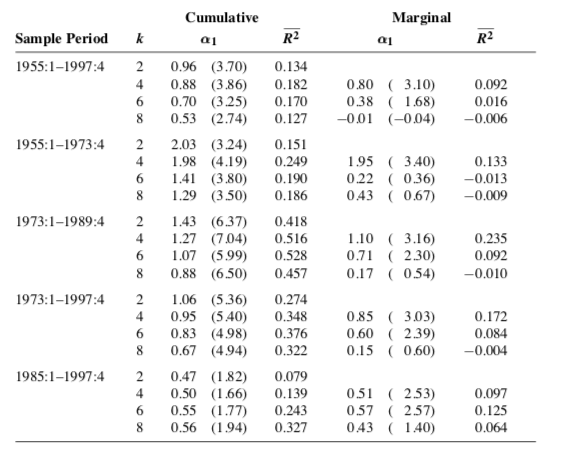}
\caption{Coefficients of Regression in Different Horizon}
\end{figure}

The coefficients are significant up to 2 years time horizon disregarding the last period. Marginal contributions are also shown in the results that suggests that the spread has explanatory power on GDP growth up to 6 months. (For 8 months marginal contribution is very close to 0) The paper concludes that the yield curve is strong indicator of future economic growth.\\

Moreover, the following table shows forecasting ability of the yield curve. In addition to the previous studies discussed, this graph includes the crisis periods to see the predictive power of the model in crisis periods. We can see from the next figure that the forecast GDP growth closely follows the actual GDP Growth Rate. Moreover, the graph also indicates that the model can also have some explanatory power on recessions. Except the recession in 1990, the forecast GDP growth can predict upcoming crises. \\

\begin{figure}[H]
\centering
\includegraphics[scale=0.8]{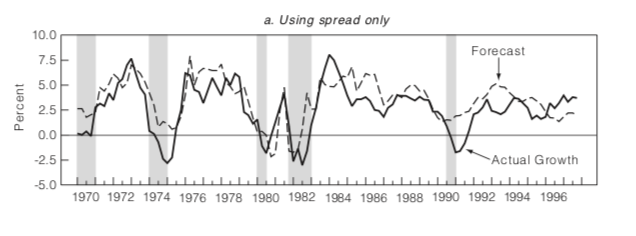}
\caption{Forecasting Performance Including Recession Periods}
\end{figure}

\subsection{Yield Curve as a Predictor of Recessions}

The studies above mostly focuses on quantitative analysis of GDP growth. However, there are many research analyzing the forecasting performance of interest rate spread on crisis periods. The question if interest rate spreads can predict the crisis is analyzed mostly by probit model as crisis is a dependent variable.\\

Estrella and Mishkin (1996) used interest rate spread to forecast the recession periods with out-of-sample examination and stated that the interest rate is more valuable tool to forecast recessions compared to other financial variables. The probit model is used to estimate probability of recession in one, two, four and six quarters in the future by using 10 years and 3 months interest rate spread and concludes that four quarters predicted probability outperform the other time horizons. The predicted probability of recessions of four quarters in the future  out of the model is shown in the following result table.\\

\begin{figure}[H]
\centering
\includegraphics[scale=0.8]{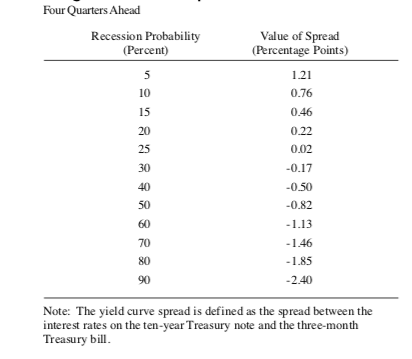}
\caption{Probability of Being in Recession}
\end{figure}

As it can be seen, if the interest rate spread increase which is a sign of normal yield curve, the predicted probability of recession decreases as we expected. Moreover, Marginal effect of interest rate spread decrease on recession probability is less as the level of interest rate decreases. This would suggest that probability of recession is increases much more as spread decreases when there is a steep normal yield curve.\\

The following figures as a result of analysis plots the one, two, four and six quarter ahead probability forecasting of yield curve and NYSE stock exchange price index on recessions to compare both performances. \\

\begin{figure}[H]
\centering
\includegraphics[scale=0.8]{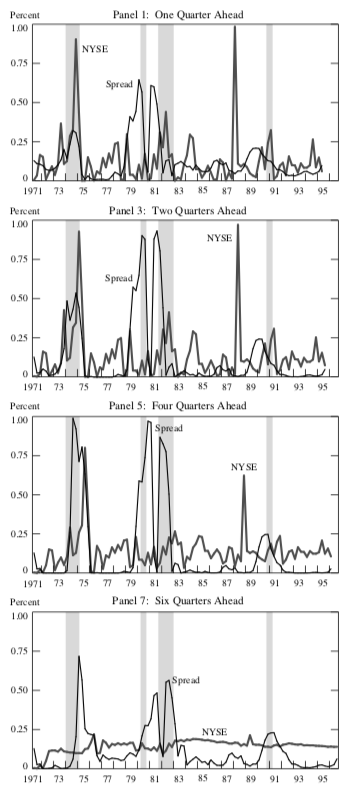}
\caption{Forecast of Recession Periods}
\end{figure}

In the graphs, there are 4 recession periods. Vertical axis stands for the predicted probability of recession. As it can be seen, the 4 quarters ahead forecast of recession gives the best result as the predicted probabilities increase significantly when there is a recession. Moreover, the forecast of yield curve outperforms the forecasting of NYSE stock exchange price index. First three recession forecasting performs quite well whereas the last recession in 1991 forecast is poor. Overall, the paper concludes that the interest rate spread has significant explanatory power on recession as it can successfully predict the recessions.\\

Furthermore, the more recent study done by Wright (2006) analyzed the forecasting performance of interest rate spread between 10 years and 3 months on recession. The quarterly data starts in 1964 and ends in 2006. They run the following probit model to estimate probability of being recession: \\
\vspace{0.5cm}
\begin{center}

\includegraphics[scale=0.7]{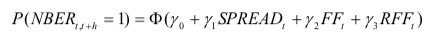}
\end{center}
\vspace{0.5cm}
where NBER stands for the announcement of Natural Bureau of Economic Research and equals to 1 if there is a recession period. SPREAD stands for the interest rate spread, FF stands for the average effective funds rate and RF stands for the real federal funds rate. Second and third independent variables is to control the level of interest rate to see the effect of spread changes alone and better forecasting performance. The study is repeated for two quarters ahead forecasting of probability of recession. The results show that the coefficient of the interest rate spread term is significantly positive as expected and has biggest explanatory power on recession as it can be seen from the following figure:  \\

\begin{figure}[H]
\centering
\includegraphics[scale=1]{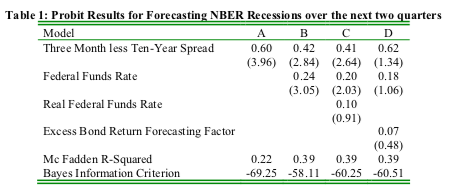}
\caption{Coefficients of Probit Model on Forecasting Recession}
\end{figure}

The predicted probability of recession out of the model can be seen from the following Figure. The red lines refers to the periods of recession out of NBER announcement. The successful performance of the model in predicting the periods of recession can be observed because the model predicts almost all of the recession as the predicted probability increases significantly on all periods of recessions whereas the predicted probability stays low in non-recession periods.\\

\begin{figure}[H]
\centering
\includegraphics[scale=0.8]{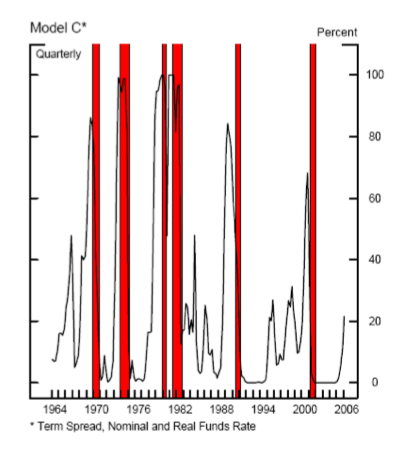}
\caption{Forecasting Recessions}
\end{figure}

\section{Yield Curve and Economic Performance: 
The Turkish Case}

In this section, I aim to analyze the relationship between the long term- short term interest rate spread and the GDP growth in Turkey, aiming to find positive relation between this two based on the findings on the literature. I used 10 and 2 years government treasury bonds to proxy interest rate spread. Positive interest rate spread stands for the normal yield curve whereas negative interest rate spread stands for the inverted yield curve. The data is limited to quarterly 2010-2017 because 10 years treasury bond data starts from 2010 and quarterly GDP growth data is available.\\

\subsection{Interest Rate Spread and GDP growth Positively Related}
In Figure 12, GDP growth and interest rate spread is plotted from 2010 to 2017. The strong  positive relationship can be observed from the graph as the movements of SPREAD closely follow the movements of GDP growth. Interpretations of the graph suggest that there is a positive relationship between Spread and GDP growth. Therefore, interest rate spread can have some explanatory power on GDP growth. Positive relationship suggests that normal steep yield curve type is a signal of a better economic performance in the future and inverted yield curve reflects the economy tending to slow down.\\

\begin{figure}[H]
\centering
\includegraphics[scale=0.8]{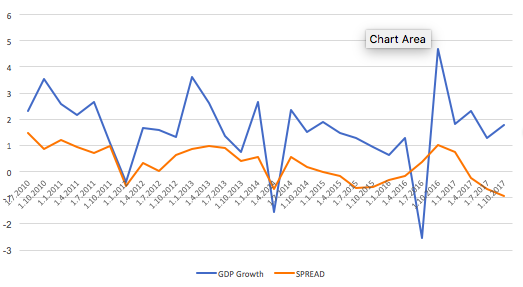}
\caption{Interest Rate Spread and GDP Growth Turkey}
\end{figure}

\subsection{Regression and Forecasting the GDP of Turkey from Yield Curve}

The predictive power of interest rate spread on GDP growth in Turkey is quite high. Following the Harvey (1989), the following regression is run to support the predictive power of interest rate spread on GDP Growth in Turkey:\\

$$Growth_{t+1}= a+b(Yield Spread)_t+\epsilon_{t}$$\\

The results can be seen in the Figure 13. The coefficient of interest rate spread is significant and equal to 1.0445 which is similar to findings in the literature for US case. This result indicates 1\% increase in the interest rate spread leads to 1.0445\% in the following period for Turkey. Therefore, outstanding results supporting positive relation supports the theoretical intuition that normal yield curve is a signal of better economic performance and inverted yield curve is interpreted as bad signal for economies.\\

\begin{figure}[H]
\centering
\includegraphics[scale=0.8]{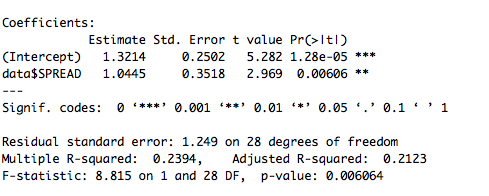}
\caption{Results}
\end{figure}

In Figure 14, forecast of Turkish GDP growth out of the model is plotted with the actual GDP. As it can be observed, the forecasted GDP growth follows closely the movements of actual GDP growth except the decrease in Q2 in 2016. This successful forecasting performance of the model reflects that interest rate spread is quite informative about the future GDP growth and economic performance.

\begin{figure}[H]
\centering
\includegraphics[scale=0.8]{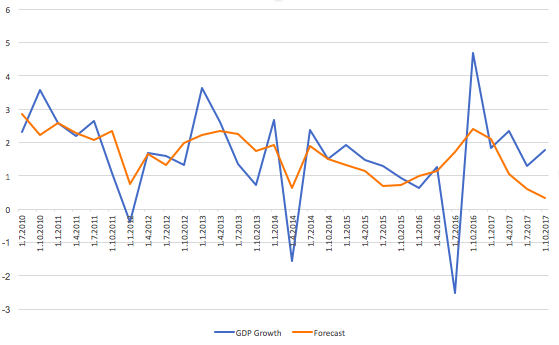}
\caption{Forecast of Turkish GDP}
\end{figure}

\section{Conclusion}

This project aims to analyze the yield curve which has information about macroeconomic dynamics of the economies. Firstly, this project look into literature to check if reading the macroeconomics from yield curve is empirically supported. Secondly, the model used in the literature is applied to Turkey with new data from 2010-2017.\\

The relevant studies mainly use interest rate spread of different maturities as a proxy of yield curve and analyze its predictive power on GDP growth. The findings supports that an increase in interest rate spread reflects a higher GDP growth. Quantitatively, the findings suggest approximately 1-to-1 relation between two, that is 1\% increase in Spread leads to 1\% increase in GDP growth next period. This significant positive relation supports the intuition behind the yield curve inferences. \\

Following the findings, the relationship was analyzed for the Turkish case by using the difference in interest rate for 10 years and 2 years as a proxy of yield curve. The findings of this project are similar to the literature and supports the positive relation in Turkey as well. 

\newpage

\textbf{REFERENCES}\\

Clark, K. (1996). "A Near-Perfect Tool for Economic Forecasting". Fortune, pp. 24-26. \\

Dotsey, Michael. (1998). "The Predictive Content of the Interest Rate Term Spread for Future Economic Growth (1998)". FRB Richmond Economic Quarterly, vol. 84, no. 3, pp. 31-51.\\

Harvey, CR.(1998). "The Real Term Structure and Consumption Growth". Journal of Financial Economics, vol. 22, no. 2, pp. 305-33. \\ 

Haubrich, J. G., \& Dombrosky, A. M. (1996). "Predicting real growth using the yield curve". Economic Review - Federal Reserve Bank of Cleveland, 32(1), 26.\\

Wright, J.H. (2006). "The yield curve and predicting recessions".  Finance and Economics Discussion Series 2006-07, Board of Governors of the Federal Reserve System (U.S.).  \\

\end{document}